\documentclass[preprint,eqsecnum,aps,nofootinbib]{revtex4}
\usepackage[dvips]{graphicx}
\usepackage{amssymb}
\usepackage{amsmath}
\usepackage{amsmath, amsthm, amssymb, mathrsfs}
\usepackage{pstricks}

\begin{document}

\title{Quantum Teleportation and Von Neumann Entropy}
\author{You Hwan Ju$^{1}$, Eylee Jung$^{1}$, Mi-Ra Hwang$^{1}$,  
D. K. Park$^{1,2}$, \\ 
Hungsoo Kim$^{2}$,
Min-Soo Kim$^{3}$, Jin-Woo Son$^{3,2}$, Sahng-Kyoon Yoo$^{4}$, S. Tamaryan$^{5}$}

\affiliation{$^1$ Department of Physics, Kyungnam University, Masan, 631-701, 
Korea \\
$^2$ The Institute of Basic Science, Kyungnam University, 
Masan, 631-701, Korea \\
$^3$ Department of Mathematics, Kyungnam University, Masan, 631-701, Korea \\ 
$^4$ Green University, Hamyang, 676-872, Korea \\
$^5$ Theory Department, Yerevan Physics Institute, 
Yerevan-36, 375036, Armenia 
}

\begin{abstract}
The single qubit quantum teleportation (sender and receiver are Alice and Bob respectively) is 
analyzed from the aspect of the quantum information theories. The various quantum entropies are
computed at each stage, which ensures the emergence of the entangled states in the intermediate
step. The mutual information $S(B:C)$ becomes non-zero before performing quantum measurement, 
which seems to be consistent to the original purpose of the quantum teleportation. It is shown that if 
the teleported state $|\psi \rangle$ is near the computational basis, the quantum measurement in
$C$-system is dominantly responsible for the joint entropy $S(A,C)$ at the final stage. If,
however, $|\psi \rangle$ is far from the computational basis, this dominant responsibility is 
moved into the quantum measurement of system $A$. A possible extension of our results are
briefly discussed.
\end{abstract}


\maketitle

\newpage
\section{Introduction}

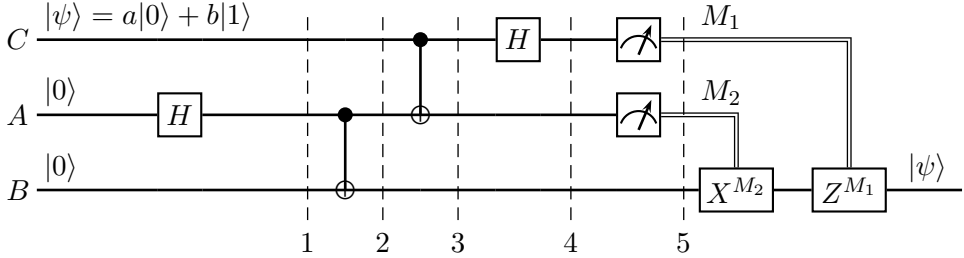
\begin{figure}
\begin{center}
\begin{pspicture}(-8,0)( 9, 3)

\rput[l](-7.0, 2.5){$ |\psi \rangle = a |0\rangle + b |1\rangle$}
\rput[l](-7.0, 1.5){$ |0\rangle $}
\rput[l](-7.0, 0.5){$ |0\rangle $}

\rput[l](-7.5, 2.2){$C$}
\rput[l](-7.5, 1.2){$A$}
\rput[l](-7.5, 0.2){$B$}

\psline[linewidth=1pt](-7.1, 2.2)(-5.5, 2.2)
\psline[linewidth=1pt](-7.1, 1.2)(-5.5, 1.2)
\psline[linewidth=1pt](-7.1, 0.2)(-5.5, 0.2)

\psline[linewidth=1pt](-5.5, 2.2)(-4.9, 2.2)
\psframe(-5.5, 0.9)(-4.9, 1.5) \rput[c](-5.2, 1.2){$H$}
\psline[linewidth=1pt](-5.5, 0.2)(-4.9, 0.2)

\psline[linewidth=1pt](-4.9, 2.2)(-3.0, 2.2)
\psline[linewidth=1pt](-4.9, 1.2)(-3.0, 1.2)
\psline[linewidth=1pt](-4.9, 0.2)(-3.0, 0.2)

\psdot[linewidth=0.05cm, dotsize=0.2cm](-3.0, 1.2)
\psline[linewidth=1pt](-3.0, 1.2)(-3.0, 0.2)
\psdots[dotstyle=oplus, dotscale=2](-3.0, 0.2)

\psline[linewidth=1pt](-3.0, 2.2)(-2.0, 2.2)
\psline[linewidth=1pt](-3.0, 1.2)(-2.0, 1.2)
\psline[linewidth=1pt](-3.0, 0.2)(-2.0, 0.2)

\psdot[linewidth=0.05cm, dotsize=0.2cm](-2.0, 2.2)
\psline[linewidth=1pt](-2.0, 2.2)(-2.0, 1.2)
\psdots[dotstyle=oplus, dotscale=2](-2.0, 1.2)

\psline[linewidth=1pt](-2.0, 2.2)(-1.0, 2.2)
\psline[linewidth=1pt](-2.0, 1.2)(-1.0, 1.2)
\psline[linewidth=1pt](-2.0, 0.2)(-1.0, 0.2)

\psframe(-1.0, 1.9)(-0.4, 2.5) \rput[c](-0.7, 2.2){$H$}
\psline[linewidth=1pt](-1.0, 1.2)(-0.4, 1.2)
\psline[linewidth=1pt](-1.0, 0.2)(-0.4, 0.2)

\psline[linewidth=1pt](-0.4, 2.2)( 0.6, 2.2)
\psline[linewidth=1pt](-0.4, 1.2)( 0.6, 1.2)
\psline[linewidth=1pt](-0.4, 0.2)( 0.6, 0.2)

\psframe( 0.6, 1.9)( 1.2, 2.5)
\psarc( 0.9, 2.0){0.25}{20}{160}
\psline[linewidth=1pt]{->}( 0.9, 2.0)( 1.07, 2.4)

\psframe( 0.6, 0.9)( 1.2, 1.5)
\psarc( 0.9, 1.0){0.25}{20}{160}
\psline[linewidth=1pt]{->}( 0.9, 1.0)( 1.07, 1.4)

\psline[linewidth=0.5pt, doubleline=true, doublesep=0.03]( 1.2, 2.2)( 3.7, 2.2)( 3.7, 0.5)
\rput[c]( 2.0, 2.5){$ M_1 $}
\psline[linewidth=0.5pt, doubleline=true, doublesep=0.03]( 1.2, 1.2)( 2.2, 1.2)( 2.2, 0.5)
\rput[c]( 2.0, 1.5){$ M_2 $}

\psline[linewidth=1pt]( 0.6, 0.2)( 1.7, 0.2)
\psframe( 1.7,-0.1)( 2.7, 0.5) \rput[c]( 2.2, 0.2){$ X^{M_2} $}
\psline[linewidth=1pt]( 2.7, 0.2)( 3.2, 0.2)
\psframe( 3.2,-0.1)( 4.2, 0.5) \rput[c]( 3.7, 0.2){$ Z^{M_1} $}

\psline[linewidth=1pt]( 4.2, 0.2)( 5.2, 0.2)
\rput[r]( 5, 0.5){$ |\psi\rangle$}

\psline[linewidth=0.5pt, linestyle=dashed](-3.5, 2.5)(-3.5,-0.2) \rput[c](-3.5, -0.5){$1$}
\psline[linewidth=0.5pt, linestyle=dashed](-2.5, 2.5)(-2.5,-0.2) \rput[c](-2.5, -0.5){$2$}
\psline[linewidth=0.5pt, linestyle=dashed](-1.5, 2.5)(-1.5,-0.2) \rput[c](-1.5, -0.5){$3$}
\psline[linewidth=0.5pt, linestyle=dashed]( 0.0, 2.5)( 0.0,-0.2) \rput[c]( 0.0, -0.5){$4$}
\psline[linewidth=0.5pt, linestyle=dashed]( 1.5, 2.5)( 1.5,-0.2) \rput[c]( 1.5, -0.5){$5$}

\end{pspicture}
\end{center}
\caption{Quantum circuit for teleporting a qubit $|\psi \rangle$. The two top lines represent Alice's 
system and the bottom line is Bob's system.}
\end{figure}

It is generally believed that Nature is governed by quantum mechanics\cite{feynman65}. Based on
this fact, Feynman suggested\cite{feynman82,feynman86} about three decades ago that 
the computer which obeys the quantum mechanical law can be made in the future. Ten years later after
Feynman suggestion P. W. Shor\cite{shor94} has shown that the computer Feynman pointed out, {\it i.e.}
quantum computer, enhances drastically the computational ability in certain mathematical problems.
Especially, Shor showed that the discrete logarithm and large integer factoring problems can be 
computed within polynomial time in the quantum computer. Recently, this factoring algorithm is 
experimentally realized in NMR\cite{vander01} and optical\cite{lu07} experiments. 
Since the efficient factoring algorithm is 
highly important in modern cryptography, Shor's factoring algorithm supports a strong motivation on
current flurry of activity in this subject. Furthermore, the recent active research on quantum 
computer provides a deep understanding in quantum mechanics, and as a result it yields a new branch of 
physics called quantum information\cite{nielsen00}.

Although the quantum computation and quantum information theories are important in the aspect of 
industrial issue, they also gives an new insight in the purely theoretical aspects. In theoretical 
physics, as well-known, one of the most important and long-standing problem is how to understand 
and formulate the quantum gravity. The most important effect of the quantum gravity, which still we 
do not fully understand, is an information loss problem in black hole physics\cite{hawk76}.
Recent development of the quantum information theories may shed light on the new avenue to understand
this highly important and fundamantal problems\cite{loss}.

In this paper we would like to examine the quantum teleportation\cite{bennett93} from the viewpoint
of the quantum entropy called von Neumann entropy. The quantum circuit for the one qubit 
teleportation is given in Fig. 1. The two top lines in Fig. 1 are Alice's system and the bottom line is
Bob's system. The main purpose of the quantum teleportation is to send the unknown quantum
state $|\psi \rangle = a |0 \rangle + b |1 \rangle$ from Alice to Bob. The state vector for each stage 
before Alice performs quantum measurement can be easily read from Fig. 1  as following:
\begin{eqnarray}
\label{states}
& &|CAB\rangle = \frac{1}{\sqrt{2}} \left[ a ( |000\rangle + |010\rangle )
+ b ( |100\rangle + |110\rangle ) \right]
\hspace{0.5cm} \mbox{(at stage 1)} \\ \nonumber
& &|CAB\rangle = \frac{1}{\sqrt{2}} \left[ a ( |000\rangle + |011\rangle )
+ b ( |100\rangle + |111\rangle ) \right]
\hspace{0.5cm} \mbox{(at stage 2)} \\ \nonumber
& &|CAB\rangle = \frac{1}{\sqrt{2}} \left[ a ( |000\rangle + |011\rangle )
+ b ( |101\rangle + |110\rangle ) \right]
\hspace{0.5cm} \mbox{(at stage 3)} \\ \nonumber
& &|CAB\rangle = \frac{1}{2} \bigg[ a ( |000\rangle + |011\rangle + 
|100\rangle + |111\rangle ) \\ \nonumber
& & \hspace{3.0cm}
+ b ( |001\rangle + |010\rangle - |101\rangle - |110\rangle ) \bigg].
\hspace{0.5cm} \mbox{(at stage 4)}
\end{eqnarray}
The quantum teleportation is possible if one uses the various pecular properties of the EPR maximally
entangled states which have no counterpart in the classical channel. In fact we can show that 
the state vector of the AB sub-system at stage ``2'' is 
\begin{equation}
\label{epr-11}
|AB \rangle = \frac{1}{\sqrt{2}} \left( |00 \rangle + |11 \rangle \right)
\end{equation}
which is one of four EPR states in two qubit system. After stage ``4'' Alice performs a quantum
measurement\footnote{The meters in Fig. 1 represent quantum measurement.} in the computational basis.
Thus the measurement outcome should be one of ($C=0, A=0$), ($C=0, A=1$), ($C=1, A=0$), and ($C=1, A=1$). 
To complete the teleportation Alice should notify Bob of the measurement result via the 
classical channel\footnote{In Fig. 1 double lines represent the classical channel.}. If Alice's 
measurement result is ($C=M_1$, $A=M_2$), Bob should operate $Z^{M_1} X^{M_2}$ to his state where
\begin{eqnarray}
\label{pauli}
X = \left(   \begin{array}{cc}
             0  &  1   \\
             1  &  0
              \end{array}    \right),
\hspace{1.0cm}
Y = \left(   \begin{array}{cc}
             0  &  -i   \\
             i  &  0
              \end{array}    \right),
\hspace{1.0cm}
Z = \left(   \begin{array}{cc}
             1  &  0   \\
             0  & -1 
              \end{array}    \right).
\end{eqnarray}
Then, as a result, Bob's state vector reduces to $|B \rangle = a |0 \rangle + b |1 \rangle$. Thus the 
state $|\psi \rangle$ is completely teleported to Bob. This is a quantum algorithm for the single
qubit quantum teleportation.

The single qubit quantum teleportation is experimentally realized in 
optics\cite{bouwmeester97,furu98}, nuclear magnetic resonance\cite{niel98} and 
ion trap experiment\cite{riebe07}. More recently, the quantum algorithm for the two qubit 
teleportation is developed\cite{yan05} via the optimal POVM measurement performed by Bob.

As stated above we would like to examine the single qubit quantum teleportation algorithm in
this paper by computing the various quantum entropies at all stages. This paper is origanized as
follows. In section II we compute the von Neumann entropies, joint entropies, relative entropies,
conditional entropies and mutual information at each stage before quantum measurement. It is shown
that the local Hadamard gate does not give any effect in quantum entropy. The mutual information
$S(B:C)$ becomes non-zero at stage ``3'' and ``4''. This means that the partial information on
$|\psi \rangle$ is moved to Bob, which is consistent with the original purpose of the teleportation. 
Several conditional entropies become negative, which indicates the appearance of the entangled 
states\cite{nielsen00}. In section III we compute the various quantum entropies at stage ``5'', where
Alice performed her measurement but still has not informed Bob of her measurement result. This situation
can be described differently as following: although Alice performs her projective measurement, for 
some reason she lost the record of her measurement result. Then it is physically reasonable to assume 
that the density operator for the joint system CAB is the averaged value produced via quantum 
measurement. It is shown that the joint and conditional entropies increase due to the projective
measurement while mutual information decreases. The physical reason for the decrease of the mutual 
information is discussed in this section. In section IV we introduce two different intermediate 
stages between stage ``4'' and ``5'' to examine the effect of the quantum measurement on quantum 
entropy. In section V a brief conclusion is given.

\section{Single-qubit Teleportation: before measurement}
In this section we would like to compute various quantities derived from von Neumann's entropy
defined
\begin{equation}
\label{entropy1}
S(\rho) \equiv -Tr \left(\rho \log \rho\right)
\end{equation}
where $\rho$ is a density operator of a given quantum system. Especially, in this section, we consider
only stage $``1''$, $``2''$, $``3''$ and $``4''$ in Fig. 1. 
The stage $``5''$(stage after quantum measurement 
performed by Alice) will be explored in the next section. Since the computational technique for
each stage is similar, we will show the calculational procedure explicitly only at stage $``3''$ and
the results for each level will be summarized in Table I, II and III. 

Using Eq.(\ref{states}) it is easy to show that the density operator $\rho^{CAB}$ for the joint
system CAB becomes
\begin{eqnarray}
\label{density1}
\rho^{CAB} \equiv |CAB\rangle \langle CAB| = 
\frac{1}{2}
\left( \begin{array}{cccccccc}
|a|^2 & 0 & 0 & |a|^2 & 0 & a b^* & a b^* & 0 \\
0 & 0 & 0 & 0 & 0 & 0 & 0 & 0 \\
0 & 0 & 0 & 0 & 0 & 0 & 0 & 0 \\
|a|^2 & 0 & 0 & |a|^2 & 0 & a b^* & a b^* & 0 \\
0 & 0 & 0 & 0 & 0 & 0 & 0 & 0 \\
a^* b & 0 & 0 & a^* b & 0 & |b|^2 & |b|^2 & 0 \\
a^* b & 0 & 0 & a^* b & 0 & |b|^2 & |b|^2 & 0 \\
0 & 0 & 0 & 0 & 0 & 0 & 0 & 0 
\end{array}
\right).
\end{eqnarray}
Since $\rho^{CAB}$ is pure state and von Neumann entropy for pure state is always 
zero\cite{nielsen00}, one 
can conclude 
\begin{equation}
\label{joint1}
S(C, A, B) = 0.
\end{equation}

Taking partial trace for Bob's system, one can directly compute $\rho^{CA}$, the density
operator for the CA joint system, whose explicit expression is 
\begin{eqnarray}
\label{density2}
\rho^{CA} = Tr_{B} \rho^{CAB} = \frac{1}{2} \left( \begin{array}{cccc}
|a|^2 & 0 & 0 & a b^* \\
0 & |a|^2 & a b^* & 0 \\
0 & a^* b & |b|^2 & 0 \\
a^* b & 0 & 0 & |b|^2
\end{array} \right).
\end{eqnarray}
Since $Tr (\rho^{CA})^2 = 1/2 \neq 1$, $\rho^{CA}$ is a mixed state. It is easy to show that 
$\rho^{CA}$ has eigenvalues $\lambda_{CA} = \{1/2, 1/2, 0, 0 \}$. Since von Neumann entropy equals to 
the classical Shannon entropy if the eigenvalues of the density operator are regarded as the probability
distribution, the quantum entropy reduces to 
\begin{equation}
\label{joint2}
S(C,A) \equiv - \sum_{i} \lambda_i \log \lambda_i = 1.
\end{equation}
By same way it is easy to show that $\rho^{CB} = \rho^{CA}$ and 
\begin{eqnarray}
\label{density3}
\rho^{AB} = Tr_C \rho^{CAB} = \frac{1}{2} \left( \begin{array}{cccc}
|a|^2 & 0 & 0 & |a|^2 \\
0 & |b|^2 & |b|^2 & 0 \\
0 & |b|^2 & |b|^2 & 0 \\
|a|^2 & 0 & 0 & |a|^2
\end{array} \right)
\end{eqnarray}
with
\begin{equation}
\label{joint3}
S(C, B) = 1, 
\hspace{1.0cm}
S(A, B) = - |a|^2 \log |a|^2 - |b|^2 \log |b|^2.
\end{equation}
Tracing out again, one can derive the density operators for the single qubit systems
\begin{eqnarray}
\label{density4}
\rho^C = \left( \begin{array}{cc}
|a|^2 & 0 \\
0 & |b|^2
\end{array} \right),
\hspace{1.0cm}
\rho^A = \rho^B = \frac{1}{2} \left( \begin{array}{cc}
1 & 0 \\
0 & 1
\end{array} \right)
\end{eqnarray}
with
\begin{equation}
\label{entropy2}
S(C) = - |a|^2 \log |a|^2 - |b|^2 \log |b|^2, 
\hspace{1.0cm}
S(A) = S(B) = 1.
\end{equation}
Eq.(\ref{density4}) implies that $\rho^A$ and $\rho^B$ are completely mixed states.

\begin{center}

\begin{tabular}{c||c|c|c||c} \hline
state & stage 1 & stage 2 & stage 3,4 & stage 5 \\ \hline \hline
$\rho^{CAB}$ & (P,P) & (P,E) & (P,E) & (M,E) \\ \hline \hline
$\rho^{CA}$ & (P,P) & (M,P) & (M,E) & (completely M,P) \\ \hline
$\rho^{CB}$ & (P,P) & (M,P) & (M,E) & (M,E) \\ \hline
$\rho^{AB}$ & (P,P) & (P,E) & (M,E) & (M,E) \\ \hline \hline
$\rho^C$ & P & P & M & completely M \\ \hline
$\rho^A$ & P & completely M & completely M & completely M \\ \hline
$\rho^B$ & P & completely M & completely M & completely M \\ \hline
\end{tabular}

\vspace{0.1cm}
Table I: The properties of the density operators at each stage. The P and M in first 
position in parenthesis denote pure and mixed respectively. The P and E in second position stand for
product and entangled respectively.
\end{center}
\vspace{1.0cm}

Using the definitions of mutual information $S(A:B) \equiv S(A) + S(B) - S(A,B)$ and conditional
entropy $S(A|B) \equiv S(A,B) - S(B)$ one can compute the various quantities summarized in Table II.

The relative entropy defined 
\begin{equation}
\label{relative1}
S(\rho || \sigma) \equiv Tr\left(\rho \log \rho \right) - Tr \left(\rho \log \sigma \right)
= - S(\rho) - Tr \left(\rho \log \sigma \right)
\end{equation}
also can be computed explicitly. Using
\begin{eqnarray}
\label{density5}
\log \rho^C = \left( \begin{array}{cc}
\log |a|^2 & 0 \\
0 & \log |b|^2
\end{array} \right),
\hspace{1.0cm}
\log \rho^A = \log \rho^B = \left( \begin{array}{cc}
-1 & 0 \\
0 & -1
\end{array} \right),
\end{eqnarray}
one can easily show
\begin{eqnarray}
\label{relative2}
& & S(\rho^C || \rho^A) = S(\rho^C || \rho^B) = 1 + |a|^2 \log |a|^2 + |b|^2 \log |b|^2
\\ \nonumber
& & S(\rho^A || \rho^C) = S(\rho^B || \rho^C) = - \log (2 |a| |b|) \\ \nonumber
& & S(\rho^A || \rho^B) = S(\rho^B || \rho^A) = 0.
\end{eqnarray}
The relative entropy for other stages is summarized at Table III. Table III shows that the relative 
entropy is always non-negative, which is known as Klein's inequality. Another point Table III indicates is
that the relative entropy sometimes becomes infinity due to $\log 0$. This is because of the non-trivial
intersection of support of $\rho$ with kernel\footnote{The support of a Hermitian operator $A$ is the 
vector space spanned by the eigenvectors of $A$ with non-zero eigenvalues. The vector space spanned
by the eigenvectors with zero eigenvalue is called kernel.} of $\sigma$. 

In order to check whether the states are entangled or not, we compare the tensor product of the 
component states with the corresponding joint state. At stage ``3'' one can show easily
\begin{eqnarray}
\label{entangled1}
& &\rho^{AB} \neq \rho^A \otimes \rho^B
\hspace{1.0cm} 
\rho^{CA} \neq \rho^C \otimes \rho^A \\ \nonumber
& &\rho^{CB} \neq \rho^C \otimes \rho^B
\hspace{1.0cm} 
\rho^{CAB} \neq \rho^C \otimes \rho^A \otimes \rho^B
\end{eqnarray}
which indicates that all joint states are entangled at this stage. The answer of the question whether
the given states are entangled or product, and mixed or pure at each stage is summarized at Table I.
In Table I (P,E) means ``pure and entangled'' and (M,P) stands for ``mixed and product''. Therefore
Table I shows that the quantum teleportation generally converts ``pure and product'' at stage ``1'' to
``mixed and entangled'' at stage ``4'' for the joint system and ``pure'' at stage ``1'' to 
``completely mixed'' at stage ``4'' for single-qubit component systems.

\begin{center}

\begin{tabular}{c||c|c|c||c} \hline
entropy & stage 1 & stage 2 & stage 3, 4 & stage 5 \\ \hline \hline
$S(A,B,C)$ & $0$ & $0$ & $0$ & $2$ \\ \hline
$S(A)$ & $0$ & $1$ & $1$ & $1$ \\ \hline
$S(B)$ & $0$ & $1$ & $1$ & $1$ \\ \hline
$S(C)$ & $0$ & $0$ & $-|a|^2 \log |a|^2 - |b|^2 \log |b|^2$ & $1$ \\ \hline
$S(A,B)$ & $0$ & $0$ & $-|a|^2 \log |a|^2 - |b|^2 \log |b|^2$ & $1 -|a|^2 \log |a|^2 - |b|^2 \log |b|^2$
\\ \hline
$S(A,C)$ & $0$ & $1$ & $1$ & $2$ \\ \hline
$S(B,C)$ & $0$ & $1$ & $1$ & $2 - \frac{1}{2} \log (1 - u^2) - 
\frac{u}{2} \log 
\frac{1 + u}{1 - u}$ \\ \hline
$S(A:B)$ & $0$ & $2$ & $2 + |a|^2 \log |a|^2 + |b|^2 \log |b|^2$ & 
$1 + |a|^2 \log |a|^2 + |b|^2 \log |b|^2$ \\ \hline
$S(B:C)$ & $0$ & $0$ & $-|a|^2 \log |a|^2 - |b|^2 \log |b|^2$ & 
$\frac{1}{2} \log (1 - u^2) + 
\frac{u}{2} \log \frac{1 + u}{1 - u}$ \\ \hline 
$S(A:C)$ & $0$ & $0$ & $-|a|^2 \log |a|^2 - |b|^2 \log |b|^2$ & $0$ \\ \hline
$S(A|B)$ & $0$ & $-1$ & $-1 -|a|^2 \log |a|^2 - |b|^2 \log |b|^2$ & 
$-|a|^2 \log |a|^2 - |b|^2 \log |b|^2$ \\ \hline
$S(A|C)$ & $0$ & $1$ & $1 + |a|^2 \log |a|^2 + |b|^2 \log |b|^2$ & $1$ \\ \hline
$S(B|C)$ & $0$ & $1$ & $1 + |a|^2 \log |a|^2 + |b|^2 \log |b|^2$ & $1 - \frac{1}{2} \log (1 - u^2)
- 
\frac{u}{2} \log
\frac{1 + u}{1 - u}$ \\ \hline
$S(B|A)$ & $0$ & $-1$ & $-1 -|a|^2 \log |a|^2 - |b|^2 \log |b|^2$ & 
$-|a|^2 \log |a|^2 - |b|^2 \log |b|^2$ \\ \hline
$S(C|A)$ & $0$ & $0$ & $0$ & $1$ \\ \hline
$S(C|B)$ & $0$ & $0$ & $0$ & $1 - \frac{1}{2} \log (1 - u^2)
-
\frac{u}{2} \log
\frac{1 + u}{1 - u}$ \\ \hline
\end{tabular}

\vspace{0.1cm}
Table II: Various quantum entropy at each stage ($u \equiv a b^* + a^* b$).

\end{center}
\vspace{1.0cm}

\begin{center}

\begin{tabular}{c||c|c|c||c} \hline
relative entropy & stage 1 & stage 2 & stage 3, 4 & stage 5 \\ \hline \hline
$S(\rho^C || \rho^A)$ & $-\frac{1}{2} (\log 0) (1 - u)$ & $1$ & $1 + |a|^2 \log |a|^2 + |b|^2 \log |b|^2$
& $0$ \\ \hline
$S(\rho^A || \rho^C)$ & $-\frac{1}{2} (\log 0) (1 - u)$ & $-1 - \frac{1}{2} (\log 0)$ & 
$-\log (2 |a| |b|)$ & $0$ \\ \hline 
$S(\rho^C || \rho^B)$ & $- (\log 0) |b|^2$ & $1$ & $1 + |a|^2 \log |a|^2 + |b|^2 \log |b|^2$ & $0$
\\ \hline
$S(\rho^B || \rho^C)$ & $- (\log 0) |b|^2$ & $-1 - \frac{1}{2} (\log 0)$ & $-\log (2 |a| |b|)$ & $0$
\\ \hline
$S(\rho^A || \rho^B)$ & $-\frac{1}{2} (\log 0)$ & $0$ & $0$ & $0$ \\ \hline
$S(\rho^B || \rho^A)$ & $-\frac{1}{2} (\log 0)$ & $0$ & $0$ & $0$ \\ \hline
\end{tabular}

\vspace{0.1cm}
Table III: Relative entropy at each stage ($u \equiv a b^* + a^* b$).

\end{center}
\vspace{1.0cm}

Now we would like to discuss Table II briefly. As is well-known, the original purpose of the quantum 
teleportation is for Alice(A) to send $\psi \rangle = a |0\rangle + b |1\rangle$ in C-system to 
Bob(B) using a Bell state $(|00\rangle + |11\rangle) / \sqrt{2}$, which is shared by Alice and 
Bob at stage ``2''. That is why the mutual information $S(B:C)$ between B and C becomes non-zero at 
stage``3'' and ``4''. Table II also shows that several conditional entropies become negative, which 
indicates the emergence of the entangled states\cite{nielsen00}. 
Another point we would like to stress is the 
fact that all joint entropies increase when stage is moved from ``4'' to ``5''. Since level ``5'' is 
a stage just after the quantum measurement performed by Alice, this fact reflects that the 
projective measurements generally increase the quantum entropy\cite{nielsen00}. 
In the next section we will
discuss how the quantities at stage ``5'' are computed.

\section{Single-qubit Teleportation: after measurement}
The stage ``5'' is just after Alice has performed the quantum measurement but just before 
Bob has learned the measurement result. We can describe the situation of the stage ``5'' differently 
as following. Firstly, Alice performed the quantum measurement at the computational besis
$\{|00\rangle, |01\rangle, |10\rangle, |11\rangle\}$ of the joint CA-system. In order to compute
the probability $P(C,A)$ for the measurement result, we need an reduced density operator
$\rho^{CA}$ at stage ``4'' which is 
\begin{eqnarray}
\label{density6}
\left( \rho^{CA} \right)_4 = \frac{1}{4}
\left(              \begin{array}{cccc}
      1  &  a b^* + a^* b  &  |a|^2 - |b|^2  &  -(a b^* - a^* b)   \\
    a b^* + a^* b  &  1  &  -(a b^* - a^* b)  & |a|^2 - |b|^2     \\
    |a|^2 - |b|^2  &  a b^* - a^* b  &  1  &  -(a b^* + a^* b)    \\
    a b^* - a^* b  &  |a|^2 - |b|^2  &  -(a b^* + a^* b)  &  1    \\
                     \end{array}
                                                    \right).
\end{eqnarray}
Thus the probability $P(0,0)$ for Alice to get $C=0$ and $A=0$ is 
\begin{equation}
\label{prob1}
P(0,0) = Tr \left[ |00\rangle \langle 00 | \left( \rho^{CA} \right)_4 \right] = \frac{1}{4}.
\end{equation}
By same way one can show easily $P(0,1) = P(1,0) = P(1,1) = 1/4$.

\begin{center}

\begin{tabular}{c|c|c|c|c} \hline
$\scriptstyle P(C.A)$ & $\scriptstyle P(0,0)=\frac{1}{4}$ & $\scriptstyle P(0,1)=\frac{1}{4}$ & 
$\scriptstyle P(1,0)=\frac{1}{4}$ & $\scriptstyle P(1,1)=\frac{1}{4}$
\\ \hline
$\scriptstyle \rho^{\scriptstyle CAB}$ & $\scriptstyle |00\rangle \langle00| 
\otimes \left( \begin{array}{cc}
\scriptstyle |a|^2 & \scriptstyle a b^* \\
\scriptstyle a^*b  & \scriptstyle |b|^2
\end{array} \right)$
& 
$\scriptstyle |01\rangle \langle01| \otimes \left( \begin{array}{cc}
|\scriptstyle b|^2 & \scriptstyle a^* b \\
\scriptstyle a b^* & \scriptstyle |a|^2
\end{array} \right)$ & 
$\scriptstyle |10\rangle \langle10| \otimes \left( \begin{array}{cc}
\scriptstyle |a|^2  & \scriptstyle -a b^* \\
\scriptstyle -a^* b & \scriptstyle |b|^2
\end{array} \right)$ & 
$\scriptstyle |11\rangle \langle11| \otimes \left( \begin{array}{cc}
\scriptstyle |b|^2  & \scriptstyle -a^* b \\
\scriptstyle -a b^* & \scriptstyle |a|^2
\end{array} \right)$ 
\\ \hline
\end{tabular}

\vspace{0.1cm}
Table IV: The probability distribution for the projective measurement performed by Alice and the 
corresponding joint density operator $\rho^{CAB}$ at stage ``5''. 

\end{center}
\vspace{1.0cm} 

In order to compute $\rho^{CAB}$ at stage ``5'' we need $\rho^{CAB}$ at stage ``4'' which is 
\begin{eqnarray}
\label{density7}
& &\rho_4^{CAB} = 
\frac{1}{4}
\Bigg[ |00\rangle \langle00| \otimes (a |0\rangle + b |1\rangle) (a^* \langle 0| + b^* \langle 1|)
                                                                                       \\  \nonumber
& &
      + |01\rangle \langle01| \otimes (b |0\rangle + a |1\rangle) (b^* \langle 0| + a^* \langle 1|)
      + |10\rangle \langle10| \otimes (a |0\rangle - b |1\rangle) (a^* \langle 0| - b^* \langle 1|)
                                                                                        \\  \nonumber
& &  \hspace{2.0cm}
      + |11\rangle \langle11| \otimes (-b |0\rangle + a |1\rangle) (-b^* \langle 0| + a^* \langle 1|)
      + \cdots  \Bigg]
\end{eqnarray} 
where $\cdots$ denotes the off-diagonal part in the joint CA-system. Thus if Alice gets 
$C = A = 0$ in the projective measurement, $\rho^{CAB}$ at stage ``5'' reduces to 
\begin{eqnarray}
\label{density8}
\rho_{0,0}^{CAB} = \frac{1}{P(0,0)} (|00\rangle \langle 00|) \rho_4^{CAB} (|00\rangle \langle 00|)
= (|00\rangle \langle 00|) \otimes \left(       \begin{array}{cc}
                                            |a|^2   &   a b^*   \\
                                            a^* b   &   |b|^2
                                                 \end{array}           \right).
\end{eqnarray}
If Alice gets different measurement results, we have, of course, different $\rho^{CAB}$ at stage ``5''.
The possible measurement results and the corresponding $\rho^{CAB}$ at stage ``5'' is summarized 
at Table IV. 

Since Alice does not inform the measurement result to Bob yet at level ``5'', the density operator
at this stage should be same with the density operator for the case that 
Alice lost, for some reason, her record of the measurement result. In the latter case it is reasonable
to conjecture $\rho^{CAB}$ as an average value as following:
\begin{eqnarray}
\label{density9}
& &\rho^{CAB} = \frac{1}{4} 
(|00\rangle \langle 00|) \otimes \left(       \begin{array}{cc}
                                            |a|^2   &   a b^*   \\
                                            a^* b   &   |b|^2
                                                 \end{array}           \right) + 
\frac{1}{4} 
(|01\rangle \langle 01|) \otimes \left(       \begin{array}{cc}
                                            |b|^2   &   a* b   \\
                                            a b^*   &   |a|^2
                                                 \end{array}           \right) 
                                                                                 \\  \nonumber
& &  
\hspace{1.0cm} +
\frac{1}{4} 
(|10\rangle \langle 10|) \otimes \left(       \begin{array}{cc}
                                            |a|^2   &   -a b^*   \\
                                            -a^* b   &   |b|^2
                                                 \end{array}           \right) +
\frac{1}{4} 
(|11\rangle \langle 11|) \otimes \left(       \begin{array}{cc}
                                            |b|^2   &   -a* b   \\
                                            -a b^*   &   |a|^2
                                                 \end{array}           \right).
\end{eqnarray}
Since $Tr(\rho^{CAB})^2 = 1/4$, it is a mixed state. Its eigenvalues are $\{1/4, 0\}$ with four-fold
degeneracies respectively, and therefore the corresponding von Neumann entropy is 
\begin{equation}
\label{entropy3-1}
S(C,A,B) = 2.
\end{equation}

Tracing out A, B, and C respectively, one can easily construct 
\begin{eqnarray}
\label{density10}
\rho^{CA} = \frac{I}{4}
\hspace{1.0cm}
\rho^{CB} = \frac{1}{4}  \left(       \begin{array}{cccc}
                                  1 & u & 0 & 0    \\
                                  u & 1 & 0 & 0    \\
                                  0 & 0 & 1 & -u   \\
                                  0 & 0 & -u & 1
                                      \end{array}             \right)
\hspace{1.0cm}
\rho^{AB} = \frac{1}{2}     \left(     \begin{array}{cccc}
                                   |a|^2  & 0 & 0 & 0     \\
                                   0 & |b|^2 & 0 & 0      \\
                                   0 & 0 & |b|^2 & 0      \\
                                   0 & 0 & 0 & |a|^2
                                         \end{array}          \right)
\end{eqnarray}
where $I$ is unit matrix and $u = a b^* + a^* b$.
It is worthwhile noting that $\rho^{CA}$ becomes completely mixed state in thi stage. This means that 
Alice's knowledge on $|\psi \rangle$ becomes completely mixed out through the quantum measurement. 
Computing the eigenvalues, one can easily compute the corresponding entropies which is explicitly 
given at Table II. Tracing out again one can also show that the density operators for all single-qubit
systems become completely mixed:
\begin{equation}
\label{density11}
\rho^C = \rho^A = \rho^B = \frac{I}{2}.
\end{equation}
This fact indicates that Bob cannot conjecture the state $|\psi \rangle$ without the classical channel
described in Fig. 1. This fact reconciles the quantum mechanics with the theory of relativity in the
faster-than-light-communication.

Table II shows that all joint entropies at stage ``5'' increase compared to those at stage ``4''.
This fact reflects the well-known fact that the projective measurement increases the quantum 
entropy. In spite of the projective measurement, however, $S(A)$ and $S(B)$ remains same at 
both stages. This is because that at stage ``2'' AB system becomes maximally entangled and therefore
the von Neumenn entropies for the component systems become maximum. Thus it is impossible to increase
the entropies although Alice performs the projective measurement between stage ``4'' and ``5''. 

Another remarkable point in Table II is that the mutual informations decrease at stage ``5'' compared
to stage ``4''. This decreasing behavior is obvious for $S(A:B)$ and $S(A:C)$ but not manifest for 
$S(B:C)$. To show that $S(B:C)$ decreases too we note 
\begin{eqnarray}
\label{mutual3}
S(B:C) = \left\{     \begin{array}{cc}
                  -r^2 \log r^2 - (1 - r^2) \log (1 - r^2)  &  \hspace{.5cm} \mbox{at stage ``4''}  \\
                  \frac{1}{2} \log (1 - u^2)  + \frac{u}{2} \log \frac{1 + u}{1 - u}  &
                                                       \hspace{.5cm}         \mbox{at stage ``5''}
                      \end{array}
                                                        \right.
\end{eqnarray}
where $u = 2 r \sqrt{1 - r^2} \cos \theta$ with $r = |a|$ and $\theta = Arg(a) - Arg(b)$. Plotting 
together one can show that $S(B:C)$ at stage ``4'' is always larger than $S(B:C)$ at stage ``5''. The
decreasing behavior of the mutual information can be explained as follows. Note that 
$S(A:B) \equiv S(A) + S(B) - S(A, B)$. Since the state of AB-system is maximally entangled at
stage ``2'', $S(A)$ and $S(B)$ become maximized before quantum measurement performed by Alice. However, 
the joint AB-system is two-qubit system, the maximum of $S(A,B)$ is two, and the joint entropy has a 
room to increase by the projective measurement. As a result, therefore, the mutual informations exhibit
decreasing behavior at stage ``5''. Finally all of the conditional entropies increase at stage ``5''. 
This can be explained too using a similar argument.

\section{Quantum Measurement Issue}
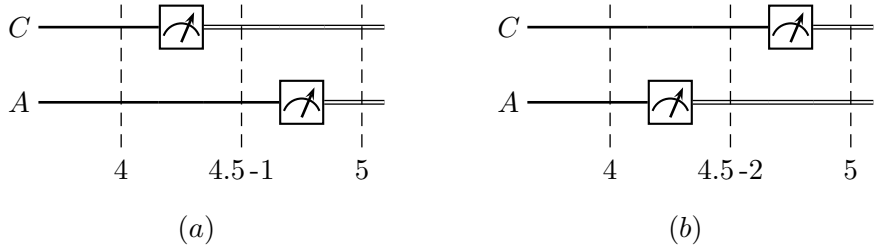
\begin{figure}
\begin{center}
\begin{pspicture}(-8,-1)(5,3)

   \rput[l](-7.5, 2.2){$C$}
   \rput[l](-7.5, 1.2){$A$}

   \psline[linewidth=1pt](-7.1, 2.2)(-5.5, 2.2)
   \psline[linewidth=1pt](-7.1, 1.2)(-5.5, 1.2)

   \psframe(-5.5, 1.9)(-4.9, 2.5)
   \psarc(-5.2, 2.0){0.25}{20}{160}
   \psline[linewidth=1pt]{->}(-5.2, 2.0)(-5.03, 2.4)
   \psline[linewidth=1pt](-5.5, 1.2)(-4.9, 1.2)

   \psline[linewidth=0.5pt, doubleline=true, doublesep=0.03](-4.9, 2.2)(-3.9, 2.2)
   \psline[linewidth=1pt](-4.9, 1.2)(-3.9, 1.2)

   \psline[linewidth=0.5pt, doubleline=true, doublesep=0.03](-3.9, 2.2)(-3.3, 2.2)
   \psframe(-3.9, 0.9)(-3.3, 1.5)
   \psarc(-3.6, 1.0){0.25}{20}{160}
   \psline[linewidth=1pt]{->}(-3.6, 1.0)(-3.43, 1.4)

   \psline[linewidth=0.5pt, doubleline=true, doublesep=0.03](-3.3, 2.2)(-2.5, 2.2)
   \psline[linewidth=0.5pt, doubleline=true, doublesep=0.03](-3.3, 1.2)(-2.5, 1.2)

   \psline[linewidth=0.5pt, linestyle=dashed](-6.0, 2.5)(-6.0, 0.6) \rput[c](-6.0, 0.3){$4$}
   \psline[linewidth=0.5pt, linestyle=dashed](-4.4, 2.5)(-4.4, 0.6) \rput[c](-4.4, 0.3){$4.5\,\mbox{-}1$}
   \psline[linewidth=0.5pt, linestyle=dashed](-2.8, 2.5)(-2.8, 0.6) \rput[c](-2.8, 0.3){$5$}

   \rput[c](-5.0,-0.5){$(a)$}


   \rput[l](-1.0, 2.2){$C$}
   \rput[l](-1.0, 1.2){$A$}

   \psline[linewidth=1pt](-0.6, 2.2)( 1.0, 2.2)
   \psline[linewidth=1pt](-0.6, 1.2)( 1.0, 1.2)

   \psline[linewidth=1pt]( 1.0, 2.2)( 1.6, 2.2)
   \psframe(1.0, 0.9)(1.6, 1.5)
   \psarc( 1.3, 1.0){0.25}{20}{160}
   \psline[linewidth=1pt]{->}( 1.3, 1.0)(1.47, 1.4)

   \psline[linewidth=1pt]( 1.6, 2.2)( 2.6, 2.2)
   \psline[linewidth=0.5pt, doubleline=true, doublesep=0.03]( 1.6, 1.2)(2.6, 1.2)

   \psframe( 2.6, 1.9)( 3.2, 2.5)
   \psarc( 2.9, 2.0){0.25}{20}{160}
   \psline[linewidth=1pt]{->}( 2.9, 2.0)(3.07, 2.4)
   \psline[linewidth=0.5pt, doubleline=true, doublesep=0.03]( 2.6, 1.2)( 3.2, 1.2)

   \psline[linewidth=0.5pt, doubleline=true, doublesep=0.03]( 3.2, 2.2)( 4.0, 2.2)
   \psline[linewidth=0.5pt, doubleline=true, doublesep=0.03]( 3.2, 1.2)( 4.0, 1.2)

   \psline[linewidth=0.5pt, linestyle=dashed]( 0.5, 2.5)( 0.5, 0.6) \rput[c]( 0.5, 0.3){$4$}
   \psline[linewidth=0.5pt, linestyle=dashed]( 2.1, 2.5)( 2.1, 0.6) \rput[c]( 2.1, 0.3){$4.5\,\mbox{-}2$}
   \psline[linewidth=0.5pt, linestyle=dashed]( 3.7, 2.5)( 3.7, 0.6) \rput[c]( 3.7, 0.3){$5$}

   \rput[c]( 1.5,-0.5){$(b)$}

\end{pspicture}
\end{center}
\caption{Two different intermediate stages between stage ``4'' and stage ``5''. These intermediate 
stages are introduced to examine the effect of quantum measurement in the change of quantum entropy.}
\end{figure}

In this section we would like to examine the issue on the effect of the quantum measurement in the 
change of quantum entropy. In order to explore this issue in detail we assume that Alice performs the
quantum measurement for C and A systems in different time as shown in Fig. 2. Thus we have two 
different stages ``4.5-1'' and ``4.5-2'' between stage ``4'' and stage ``5''. 

Now we consider stage ``4.5-1'', where the projective measurement is performed at the 
computational basis $\{ (|0\rangle \langle 0|)_C, (|1\rangle \langle 1|)_C \}$. In order to compute
the probability for the measurement results $C=0$ or $C=1$ we need $\rho^{C}$ at stage ``4'', which is 
\begin{equation}
\label{density41}
(\rho^C)_4 = \frac{1}{2} \left[ |0\rangle \langle 0| + (|a|^2 - |b|^2) |0\rangle \langle 1| 
+ (|a|^2 - |b|^2) |1\rangle \langle 0| + |1\rangle \langle 1| \right].
\end{equation}
Thus the probabilities become
\begin{eqnarray}
\label{prob41}
& &P(C=0) = Tr[ |0\rangle \langle 0| (\rho^C)_4] = \frac{1}{2}     \\   \nonumber
& &P(C=1) = Tr[ |1\rangle \langle 1| (\rho^C)_4] = \frac{1}{2}.
\end{eqnarray}
When the quantum measurement yields C=0, the density operator for CAB joint system becomes 
\begin{equation}
\label{density42}
\rho_0^{CAB} = \frac{1}{P(C=0)}
\left[ (|0\rangle \langle 0|)_C \rho_4^{CAB} (|0\rangle \langle 0|)_C \right]
\end{equation}
where $\rho_4^{CAB}$ is density operator for CAB system at stage ``4''. By same way it is 
straightforward to compute $\rho_1^{CAB}$, {\it i.e.} the density operator for CAB system when 
Alice gets C=1, which is 
\begin{equation}
\label{density43}
\rho_1^{CAB} = \frac{1}{P(C=1)}
\left[ (|1\rangle \langle 1|)_C \rho_4^{CAB} (|1\rangle \langle 1|)_C \right].
\end{equation}
If Alice lost the record of her measurement result, the density operator reduces to its 
expectation value
\begin{equation}
\label{density44}
\rho^{CAB} = P(C=0) \rho_0^{CAB} + P(C=1) \rho_1^{CAB}.
\end{equation}

Once the density operator for total system is obtained, it is easy to compute the density operator
for the sub-systems by making use of the partial trace appropriately. Then one can compute the 
von Neumann entropies easily. Of course similar computational procedure can be applied to 
compute the quantum entropy at stage ``4.5-2''. The various quantum entropies at
stage ``4.5-1'' and ``4.5-2'' are summarized at Table V when Alice lost her record of the 
measurement result.

\begin{center}

\begin{tabular}{c|c|c} \hline
        &  stage ``4.5-1''   &   stage ``4.5-2''   \\  \hline
$S(A,B,C)$  & $1$   &   $1$                        \\  \hline
$S(A)$   &  $1$     &   $1$                        \\  \hline
$S(B)$   &  $1$     &   $1$                        \\  \hline
$S(C)$   &  $1$    & $-|a|^2 \log |a|^2 - |b|^2 \log |b|^2$   \\   \hline
$S(A,B)$  &  $-|a|^2 \log |a|^2 - |b|^2 \log |b|^2$  & $1-|a|^2 \log |a|^2 - |b|^2 \log |b|^2$ \\  \hline
$S(A,C)$ & $2 - \frac{1}{2} \log (1 - u^2) - \frac{u}{2} \log \frac{1 + u}{1 - u}$  &
                                             $1-|a|^2 \log |a|^2 - |b|^2 \log |b|^2$ \\  \hline
$S(B,C)$ & $2 - \frac{1}{2} \log (1 - u^2) - \frac{u}{2} \log \frac{1 + u}{1 - u}$  & $1$  \\ \hline
\end{tabular}

\vspace{0.1cm}
Table V:Various quantum entropy at stage ``4.5-1'' and stage ``4.5-2'' ($u \equiv a b^* + a^* b$). 

\end{center}
\vspace{1.0cm}

The quantum entropies in the intermediate stages have several properties. For example, the quantum
state in the C-system becomes completely mixed at stage ``4.5-1'', which is same with that of 
stage ``5''. However, at stage ``4.5-2'' $S(C)$ is same with that of stage ``4''. This fact indicates
that measuring C performed by Alice is responsible for the change of $S(C)$ between stage ``4'' and 
``5''. Same and converse situations occur in $S(B,C)$ and $S(A,B)$ respectively. The only one which has 
non-trivial value in the intermediate stage is $S(A,C)$. Of course $S(A,C)$ in the intermediate 
stages ``4.5-1'' and ``4.5-2'' are between $1$ and $2$, where the former is $S(A,C)$ at stage ``4'' while
the latter is at stage ``5''. Defining $r = |a|$ and $\theta = Arg(a) - Arg(b)$ again, one can plot 
$S(A,C)$ which is given in Fig. 3.

\begin{figure}[ht!]
\begin{center}
\includegraphics[height=10cm]{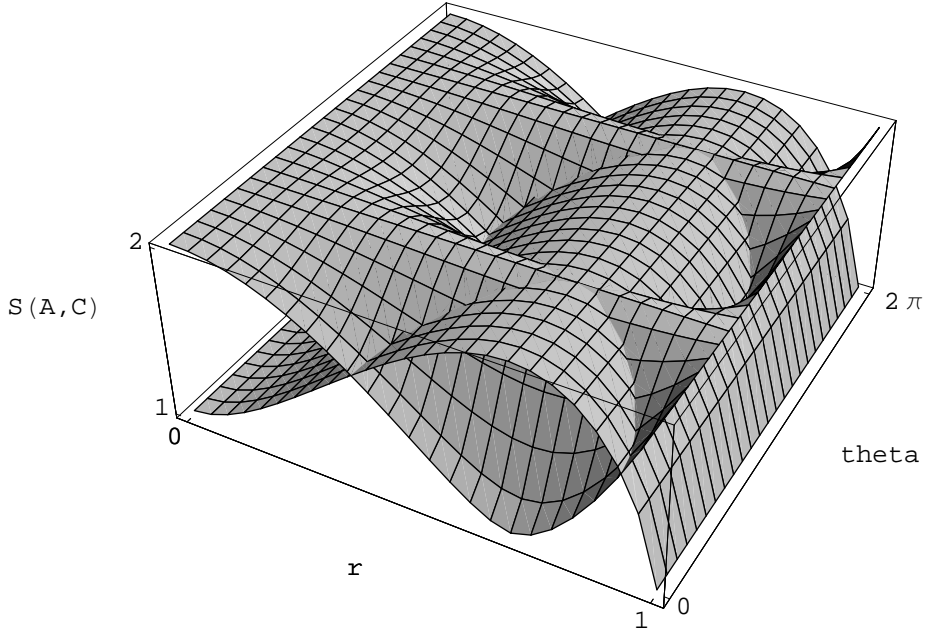}
\caption[fig3]{Plot of joint entropy $S(A,C)$ at intermediate stages ``4.5-1'' and ``4.5-2'' with 
varying $0 \leq r \equiv |a| \leq 1$ and $0 \leq \theta \equiv Arg(a) - Arg(b) \leq 2 \pi$. 
This figure indicates that if $|\psi \rangle$ is close to the computational basis ($r \sim 0$ or
$r \sim 1$), $S(A,C)$ in the stage ``4.5-1'' is much larger than that in the stage ``4.5-2'' while 
converse situation occurs when $|\psi \rangle$ is far from the computational basis.}
\end{center}
\end{figure}

Fig. 3 indicates that in the small $r$ and large $r$ region $S(A,C)$ in the stage ``4.5-1'' is much larger
than that in stage ``4.5-2''. However, in the intermediate range $S(A,C)$ in the stage ``4.5-2'' becomes
much larger. This means that if $|\psi \rangle$ is close to the computational basis, measuring C is 
dominantly responsible for $S(A,C)$ at stage ``5''. If, however, $|\psi \rangle$ is far from 
the computational basis, this responsibility is changed into the measurment of system A.

\section{conclusion}
In this paper we have analyzed the single qubit quantum teleportation by computing the various quantum
entropies at each stage of Fig. 1. Before quantum measurement performed by Alice between stage ``4''
and stage ``5'', the von Neumann entropies, conditional entropies, relative entropies and mutual
information are summarized in Table II and III. Table III shows that the relative entropy is always
non-negative, which is well-known as Klein's inequality. Therefore, the relative entropy can be 
regarded as a measure for distance between two different quantum states like trace distance or 
fidelity. Some relative entropies become infinity, which indicates the non-trivial intersection of 
the support of one quantum state with kernel of the other quantum state. Table II shows that the mutual
information $S(B:C)$ becomes non-negative at stage ``3'' and ``4''. This means that the partial 
information on $|\psi \rangle$ is transmitted to Bob, consistent with the original purpose of 
the quantum teleportation. Table II also shows that some conditional entropies become negative when 
the corresponding joint systems are in pure states. This fact indicates that the component systems are
entangled. Of course there are many entangled states in sub-systems when the joint system is not
in pure state. The properties of entangled or product, and pure or mixed for all systems are 
summarized in Table I.

At stage ``5'', where Alice performs the projective measurement but she has not yet informed of the 
measurement result to Bob through classical channel, the state for the joint system CAB can be chosen
as an average expectation value. In this case the state of total system CAB becomes mixed and 
entangled. The various quantum entropies are summarized at Table II and Table III. Table II shows that 
the joint and conditional entropies increase due to the projective measurement while the mutual 
information decreases. The reason for the decrease of the mutual information is discussed in section III.

Finally we have introduced two different intermediate stages ``4.5-1'' and ``4.5-2'' between 
stage ``4'' and ``5'' in Fig. 2 to examine the effect of the quantum measurement in the quantum entropies.
The quantum entropies in these intermediate stages are summarized in Table V. Table V shows that all
entropies except $S(A,C)$ are either one of corresponding entropies at stage ``4'' or stage ``5''. The
joint entropy $S(A,C)$ at stage ``4.5-1'' and ``4.5-2'' are plotted in Fig. 3. From this figure we can 
understand that if $|\psi\rangle$ is close to the computational basis, measuring $C$ is dominantly 
responsible for $S(A,C)$ at stage ``5'' while this dominant responsibility is changed into the 
measurement of $A$ if $|\psi\rangle$ is far from the computational basis.

It is of interest to extend our results to the quantum algorithm for the multi-qubit quantum 
teleportation. Also it seems to be interest to analyze the Shor's factoring algorithm\cite{shor94} and 
Grover's search algorithm\cite{grover96,grover97} from the aspect of the quantum information 
theories. We hope to visit these issues in the near future.

\vspace{1cm}

{\bf Acknowledgement}: 
This work was supported by the Kyungnam University
Research Fund, 2006.

\end{document}